\documentclass[reprint,amsmath,amssymb,aps,prl,showpacs]{revtex4-1}
\usepackage{amsmath,amssymb,amsfonts}
\usepackage{graphicx,bm}
\usepackage{slashed}
\usepackage{verbatim}
\usepackage[usenames]{color}
\usepackage{hyperref}
\usepackage{subcaption}

\usepackage[percent]{overpic}

\DeclareGraphicsExtensions{.jpg,.JPG,.jpeg,.pdf,.png,.eps}

\def\be{\begin{equation}}
\def\ee{\end{equation}}
\def\bea{\begin{eqnarray}}
\def\eea{\end{eqnarray}}
\newcommand{\ba}{\begin{aligned}}
\newcommand{\ea}{\end{aligned}}
\newcommand{\baa}{\begin{array}}
\newcommand{\eaa}{\end{array}}
\def\ben{\begin{enumerate}}
\def\een{\end{enumerate}}

\newcommand\fverb{\setbox\pippobox=\hbox\bgroup\verb}
\newcommand\fverbdo{\egroup\medskip\noindent\fbox{\unhbox\pippobox}\ }
\newcommand\fverbit{\egroup\item[\fbox{\unhbox\pippobox}]}

\newcommand{\bear}{\begin{eqnarray}}
\newcommand{\eear}{\end{eqnarray}}

\newbox\pippobox

\relax

\def\6{\partial}

\def\a{\alpha}

\def\C0{{\bf C_0}}
\def\Y0{{\bf Y_0}}
\def\G0{{\bf G_0}}

\def\sq
\def\a{\alpha}

\def\bz{\begin{itemize}}
\def\ez{\end{itemize}}
\def\bn{\begin{enumerate}}
\def\en{\end{enumerate}}

\def\ben{\begin{enumerate}}
\def\een{\end{enumerate}}

\def\a{\alpha}

\def\6{\partial}

\def\be{\begin{equation}}
\def\ee{\end{equation}}
\def\bea{\begin{eqnarray}}
\def\eea{\end{eqnarray}}
\def\bz{\begin{itemize}}
\def\ez{\end{itemize}}

\begin{document}

\title{Holographic Equilibration of Nonrelativistic Plasmas}
\author{Umut G\"ursoy}
\author{Aron Jansen}
\author{Watse Sybesma}
\author{Stefan Vandoren}
\affiliation{Institute for Theoretical Physics and Center for Extreme Matter and Emergent Phenomena, Utrecht University, Leuvenlaan 4, 3584 CE Utrecht, The Netherlands}
\date{\today}

\begin{abstract} 
We study far-from-equilibrium physics of strongly interacting plasmas at criticality and zero charge density for a wide range of dynamical scaling exponents $z$ in $d$ dimensions using holographic methods. 
In particular, we consider homogeneous isotropization of asymptotically Lifshitz black branes with full backreaction. 
We find stable evolution and equilibration times that exhibit small dependence of $z$ and are of the order of the inverse temperature. 
Performing a quasinormal mode analysis we find a corresponding narrow range of relaxation times, fully characterized by the fraction $z/(d-1)$. For $z\!\geq\! d\!-\!1$, equilibration is overdamped, whereas for $z\!<\!d\!-\!1$ we find oscillatory behavior.
\end{abstract}

\maketitle

\paragraph{Introduction.---}

Quantum criticality has been a focus of interest both in theoretical and experimental physics over the past few decades. It is believed to be a key ingredient in the solution to the various yet unsolved problems such as the high $T_c$ superconductivity \cite{sachdev2007quantum}. In particular, the dynamics of a system near a {\em continuous} quantum phase transition is governed by a universal, scale invariant theory characterized by the dimensionality $d$, the dynamical scaling exponent $z$ and the various other critical exponents that are independent of the microscopic Hamiltonian of the system. In these systems, the characteristic energy scale $\Delta$, such as the gap separating the first excited state from the ground state, vanishes as the correlation length $\xi$ diverges as $\Delta \sim \xi^{-z}$. For instance, $z=1$ occurs at the touching points of the band structure of mono-layer graphene and $z=2$ can describe the case of bilayer graphene, see e.g. \cite{Hartnoll:2009sz,Neto:2009aa,Killi:2011aa}.

The existence of a quantum critical point at vanishing temperature determines the behavior of observables also at finite temperature, and even beyond the thermal equilibrium, in the so called {\em quantum critical region} of the parameter space. In fact, a basic way to characterize this quantum critical region is to consider the response of the system to a {\em small disturbance}, determined by the equilibration time $\tau_{eq}$ \footnote{We use the terminology {\em relaxation time} for equilibration after small perturbations. Equilibration also occurs after large perturbations, and the isotropization we study in this paper is an example of this. In that case, we speak of {\it isotropization time}.}. The quantum critical region corresponds to short relaxation times $\tau_{eq} \sim 1/T$ \footnote{We work in units where $\hbar = k_B = c =1$.}, whereas local equilibrium is reached much more slowly as $\tau_{eq} \gg 1/T$ outside the quantum critical region \cite{sachdev2007quantum}. 

In this paper, we want to go one step further and ask the question, what happens when such a quantum critical system is taken completely out of equilibrium, when the perturbation is not small, but of the same order as the Hamiltonian. We answer this question partially in the particular situation when the collective excitations of the system are characterized by global, hydrodynamic quantities such as energy and pressure gradients at any time during the evolution. In this case, an example of such a {\em large perturbation} would be to consider a homogenous, isotropic system with average pressure $P$, and create an initial anisotropy in one direction, say $\Delta P_x$, that is  of the same order as $P$. The question then is how to characterize the evolution of this system towards equilibrium. 

We will investigate equilibration processes in strongly interacting systems that can be modeled by holography \cite{Maldacena:1997re,Gubser:1998bc,Witten:1998qj}. There is, by now, a substantial amount of work in the literature that goes by the name {\em holographic thermalization}, concerning this problem in the case of {\em relativistic} scaling, $z=1$, following the seminal work of \cite{Chesler:2008hg}. The problem of equilibration is mapped onto the evolution of a black brane geometry in the dual gravitational description. Thus, we obtain the fully nonlinear evolution of the black brane starting from the aforementioned initial conditions that correspond to anisotropy in pressure and we determine the isotropization of the system in time. In accordance with the earlier results for $z=1$, e.g. \cite{Fuini:2015hba,Ishii:2015gia}, we find that the system equilibrates quite rapidly, with isotropization times of the order $\sim 1/T$.  At later times, close to global thermal equilibrium, the evolution of the system is characterized by the {\em quasinormal modes}  of the black brane \cite{Horowitz:1999jd}. In particular the relaxation time $\tau$ above is related to the lowest lying quasinormal frequency as $\tau = - 1/\text{Im}\, \omega_0$. We find that the relaxation times are determined by the ratio $z/(d-1)$ and fall in a narrow range. This agrees with our results of the nonlinear evolution.
\paragraph{Gravitational model.---}
The holographic description of field theories with Lifshitz scaling at criticality and zero temperature was initiated in \cite{kachru:2008yh}. 
To describe nonrelativistic plasmas holographically in the critical region at non-zero temperature,  black brane solutions with Lifshitz asymptotics can be used \cite{Taylor:2008tg}. 
The Hawking temperature of the black brane corresponds, via holography, to the temperature of the dual field theory. The action for this model is given by
\begin{equation} 
\!\!S =   \frac{1 }{16\pi G}\int \!\!d^{d+1}x \sqrt{-g}\left[R \!-\! \Lambda\! -\! \frac{1}{2} (\partial \phi)^2\! - \!\frac{e^{\lambda \phi}}{4} F^2  \right] \, , 
\end{equation}
where $\Lambda =- (d+z-1)(d+z-2) $ is the cosmological constant, $\lambda = - \sqrt{\frac{2(d-1)}{z-1}}$ and $z$ is the dynamical scaling exponent, which is bounded by the null energy condition to be $z \geq 1$ \cite{Hoyos:2010at}. 
The scalar field $\phi$ and the gauge field $F = d A$ are needed to support the Lifshitz geometry.

We consider an anisotropic but homogeneous system, which we describe using the following Ansatz \footnote{Notice that this ansatz still has the gauge symmetry $r \rightarrow \left( r^z + \eta(t) \right)^{1/z}$ which leaves the form invariant. In this letter $\eta = 0$ is chosen.}
\begin{equation}\label{eq:ansatz}\begin{split}
ds^2 &= - f(t,r) dt^2 + 2 r^{z-1} dt dr + \\
&+ S(t,r)^2 \left[ e^{(d-2) B(t,r)} d x_1^2 + e^{-B(t,r)} d\vec{x}^2_{d-2} \right] \, , \\
A &= a(r,t) dt \, , \, \,\,\,\, \quad \phi = \phi(r,t) \, .
\end{split}\end{equation}
We take the gauge $A_r = 0$, which is essential to obtain the nested form of the equations (\ref{eq:dynamical}). The function $B(t,r)$ expresses the anisotropy of the black brane. 
The boundary, where the plasma lives, is at $r \rightarrow \infty$ and the horizon is denoted by $r_H$. 
From the viewpoint of holography this setup is dual to a nonrelativistic plasma with a pressure difference between the longitudinal direction $x_{1}$ and transversal directions $\vec{x}_{d-2}$.

By solving the equations of motion near the boundary, imposing Lifshitz asymptotics, we obtain 
\begin{subequations}
\begin{eqnarray}
\label{eq:nbexpansion}
f(r,t) &=& r^{2z} + \mathcal{E}r^{z+1-d} - \frac{5}{8} \frac{\mathcal{P}(t)^2}{r^{2 d - 2} } +  ... \, , \\ 
S(r,t) &=& r - n_S \frac{\mathcal{P}(t)^2}{r^{2(d+z)-3} } + ...  \, , \\
\!\phi(r,t) &=& \!\phi_0\!  \log \! \frac{ r^{2(d-1)}}{2(z\!-\!1)(z\!+\!d\!-\!1)}\! - \!\frac{2n_{S}\mathcal{P}(t)^2}{r^{2(d+z-1)} } \!+\! ... \, ,  \\
a^\prime(r,t) &=&  r^{z+d - 2} - (d-1)n_{S} \frac{\mathcal{P}(t)^2}{r^{d+z} } +  ...\, , \\ 
B(r,t) &=&   \frac{\mathcal{P}(t)}{r^{d+z-1} }+  \frac{1}{z} \frac{\partial_t\mathcal{P}(t) }{r^{d+2z-1}} +...\, .
\end{eqnarray}\end{subequations}
Where $n_S=((d \!-\! 2)/8) (d \!+ \!z\! -\! 1)/(2 (d\! -\! 1) \!+\! z)$ and $\phi_0 = \sqrt{\frac{1}{2}(z-1)/(d-1)}$. 
In contrast to \cite{Chesler:2008hg} we do not quench the system, but consider the equilibration of an out of equilibrium state, so do not turn on a source for $B$.

In this expansion there are two free coefficients $\mathcal{E}$ and $\mathcal{P}$. 
The $\mathcal{E}$ is the normalizable mode of $f$ and it is proportional to the energy, which is required to be constant by the equations of motion.
The function $\mathcal{P}$ is the normalizable mode of $B$ and will be related to the pressure difference. For $\mathcal{P}=0$ we recover the static black brane solution with $\mathcal{E}=-r_H^{d+z+1}$ \cite{Taylor:2008tg}. The Hawking temperature is given by $T_0 = \frac{d+z-1}{4\pi} r_H^z$.
We stress that the scalar and gauge field do not have independent modes, they are completely determined by the metric and do not have any intrinsic dynamics.

To obtain vacuum expectation values, we need the counterterm action on the boundary. For our setup this can be obtained by generalizing e.g. the analysis of \cite{Kiritsis:2015doa} to arbitrary dimensions, 
\begin{equation}\label{eq:ct}
\!\!S_{ct} = \frac{1}{8\pi G}\!\!\int \!\!d^d x \sqrt{-\gamma}  \left[ z\! -\!2 d\! -\! 3 \!+\! \frac{d+z-1}{2}e^{\lambda \phi}A^2    \right] \, .
\end{equation}
Notice that it breaks gauge invariance, but this is not an issue since the gauge field is not normalizable and not used to induce a chemical potential on the boundary \footnote{Throughout this paper we work at zero charge density. To introduce a chemical potential, one may use the black brane solution of \cite{Tarrio:2011de}}. For $z=1$ there is no gauge field, so the first term is enough.

Following \cite{Kiritsis:2015doa} this yields a boundary energy momentum tensor in the coordinate basis $(t,x_{1},\vec{x}_{d-2})$, 
\begin{equation}
T_{\mu\nu} = \frac{N^2}{2\pi^2}\text{diag}\left(E,P_L,P_T,...,P_T\right) \, ,
\end{equation}
where $E = - \frac{d-1}{2} \mathcal{E}$ and $\Delta P \equiv P_L - P_T = \frac{(d-1)(d+z-1)}{2} \mathcal{P}(t)$. 
It satisfies the Ward identity $z E = P_L + (d-2) P_T$. In equilibrium the pressure $P_{0}=-\frac{z}{2}\mathcal{E}$. Furthermore, we abbreviated $N^2/(2\pi^2)=1/(8\pi G)$.

\paragraph{Numerical Methods.---}
The numerical method we use to obtain solutions is an adaptation of \cite{Chesler:2008hg} to asymptotically Lifshitz spacetimes. Using the ansatz in Eq. (\ref{eq:ansatz}) and working with null derivatives $h^\prime \equiv \partial_r h$ and $\dot{h} \equiv \partial_t h + \frac{1}{2} r^{1-z} f \partial_r h$, the equations of motion can be put in a nested structure of linear ODEs,
\begin{subequations}
\begin{eqnarray}
\label{Seq}
0 &=& S^{\prime\prime} +\frac{1-z}{r} S^\prime +  \frac{1}{2(d-1)} S  {\phi^\prime}^2+\frac{d-2}{4} S {B^\prime}^2 \, ,
 \\ \label{aeq}
0 &=& a^{\prime\prime} + a^\prime \left( \frac{1-z}{r}+(d-1) \frac{S^\prime}{S}  + \lambda \phi^\prime \right) \, , \\
\label{Sdoteq}
0 &=& (\dot{S})^\prime +(d-2) \frac{S^\prime}{S} \dot{S} + \frac{\Lambda \, r^{z-1} S }{2(d-1)}   +\frac{r^{1-z} Se^{\lambda \phi}}{4(d-1)} \,  {a^\prime}^2 \, , \\
\label{Bdoteq}
0 &=&  (\dot{B})^\prime + \frac{d-1}{2} \frac{S^\prime}{S} \dot{B} + \frac{d-1}{2} \frac{\dot{S}}{S}  B^\prime  \, , \\
\label{phidoteq}
0 &=& (\dot{\phi})^\prime +\frac{d-1}{2} \frac{S^\prime}{S} \dot{\phi} +\frac{d-1}{2} \frac{\dot{S}}{S} \phi^\prime+ \frac{\lambda}{4} r^{1-z} e^{\lambda \phi} {a^\prime}^2  \, , \\
\label{feq}
0 &=& f^{\prime\prime} +\frac{1-z}{r} f^\prime -2(d-1)(d-2) r^{z-1}\frac{\dot{S} S^\prime}{S^2}+ r^{z-1} \dot{\phi} \phi^\prime \\
&+& \!\frac{(d\!-\!1)(d\!-\!2)}{2} r^{z-1}\!\dot{B} B^\prime-\frac{d\!-\!3}{d\!-\!1} \Lambda \, r^{2z-2}   -\frac{1}{2} \frac{3d\!-\!5}{d-1} e^{\lambda \phi} {a^\prime}^2\, ,  \notag \\
\label{adoteq}
0 &=& \dot{\left(a^\prime\right)} + a^\prime \left( (d-1) \frac{\dot{S}}{S}  + \lambda \dot{\phi}  - \frac{z-1}{2}r^{-z}  f \right) \, , \\
\label{constreq}
0 &=& \ddot{S} + \frac{1}{2(d-1)}  S \dot{\phi}^2 - \frac{1}{2} r^{1-z} \dot{S} f^\prime + \frac{d-2}{4} S \dot{B}^2  \, . 
\end{eqnarray}
\label{eq:dynamical}
\end{subequations}

Given initial profiles $B(t\!=\!0,r)$ and $\phi(t\!=\!0,r)$, the first equation is an ODE for $S$. Having solved this, the second becomes an ODE for $a^\prime$, and so we solve for the full geometry at $t=0$ by solving linear ODE's. In these equations we consider $\dot{S}$ independent of $S$, and similarly for the other functions.
Numerically, we use pseudospectral methods \cite{boyd2001chebyshev} to solve these equations. All plots were generated with a grid of 40 points. After solving one timestep, we can use $B$, $\dot{B}$ and $f$ to find $\partial_t B$ from the definition of the dot. We use a 4th order Adams Bashforth stepper to evolve this to the next timestep. We do the same for $\phi$, and then start the procedure again \footnote{There is a small subtlety here which does not arise in AdS. We have to specify initial profiles for both $B$ and $\phi$, but $\phi$ does not have any independent dynamics. Therefore we must be careful to choose self-consistent initial profiles. We use the near boundary expansion to order $3(d+z-1)$ to achieve this.}. 

For the numerics we change the radial coordinate to $u \!=\! r^{-z}$. One computes the event horizon by solving the equation $\frac{\partial_{t}u_{H}(t)}{u_H(t)^2}\! =\! - \frac{z}{2} f(u_H(t),t)$, arising from $ds^2\!\!=\!0$, with the boundary condition at late times $f(u_H(t),t)|_{t \rightarrow \infty} \!= \!0$. 
We also have an apparent horizon, defined as the location of the largest trapped surface, given by $\dot{S} (u_{AH}(t),t) \!=\! 0$. 
Using the Lifshitz scaling symmetry we set $\mathcal{E}\!=\!-1$ so that at equilibrium $r_H\! =\! u_H \!= \!1$. 

\paragraph{Results.---}

\begin{figure}[!tbp]
   {
     \begin{subfigure}[t]{0.23\textwidth}
        \centering
        \begin{overpic}[width=1\textwidth]{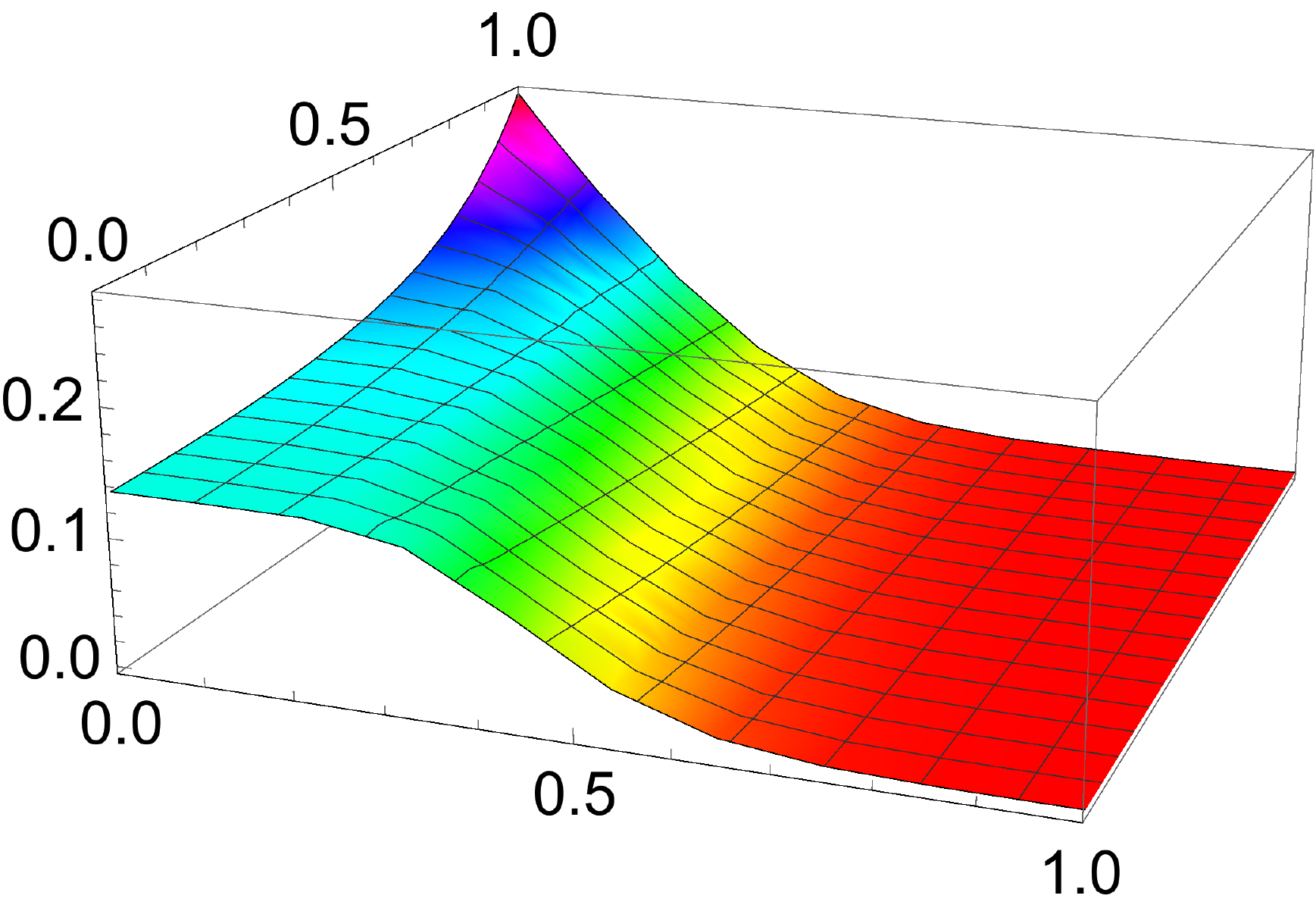}
        
	\put (52,0) {$\displaystyle T_0\,t$}
        \put (-15,30) {$\displaystyle \frac{B}{u^{5/2}}$}
        \put (10,65) {$\displaystyle u/u_{H} $}
        
        \end{overpic}
        \caption{}\label{fig:3dplot}
    \end{subfigure}
    \hfill
    \begin{subfigure}[t]{0.23\textwidth}
        \centering
        \includegraphics[width=\textwidth]{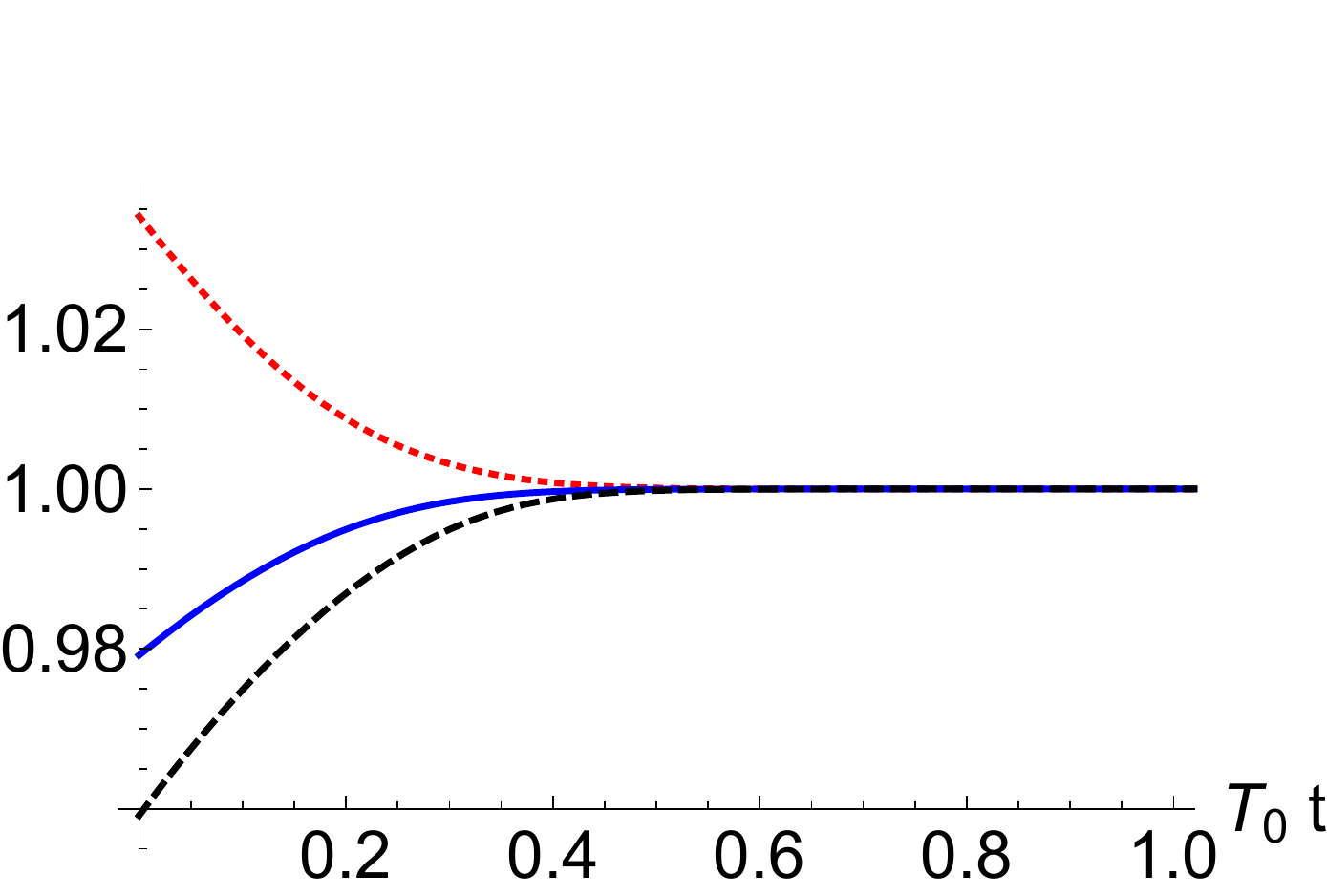}
        \caption{}\label{fig:entropy}
    \end{subfigure}
    }
    
    \captionsetup{justification=raggedright}
\caption{(a) nonlinear evolution of $B(u,t)$ for $d = 4$, $z=2$. The colors indicate equal height. 
(b): time evolution of the temperature $T/T_0$ (red, dotted), event horizon area $S/S_0$ (blue), event horizon (blue) and apparent horizon area $S_{AH}/S_{AH,0}$ (black, dashed), as a function of $T_0 t$, where $T_0$ is the equilibrium temperature, for $d=4$, $z=2$. }
\end{figure}

We analyse the nonlinear evolution for the cases $3 \leq d \leq 5$ and $1 \leq z \leq 4$. 
For the initial profiles we take the pressure to be $\frac{\Delta P(t=0)}{P_0} = 1$. This defines $\mathcal{P}$ (we set time derivatives to 0) and to obtain consistent profiles we plug this in the near boundary expansion (\ref{eq:nbexpansion}).

We highlight the case $d=4$ and $z=2$, which is physically the most interesting. 
In Figure \ref{fig:3dplot} the evolution of the anisotropy function $B$ over the whole bulk spacetime is plotted. 
Next to it is a plot showing the effects of backreaction in the same evolution. 
We plot the area density of the event horizon and the apparent horizon in Figure \ref{fig:entropy}, in \cite{Hubeny:2007xt} it is argued that the latter corresponds to the entropy, which strictly increases here. 
These are all consistency checks of our numerics \footnote{As a final check on the numerics, the constraint equation Eq. (\ref{constreq}) is small and decreases further with increasing precision.}.
We also see that the temperature decreases in time. 
In all cases considered, the above consistency checks are valid and Figure \ref{fig:entropy} looks qualitatively the same.

\begin{figure}[h]
\begin{center}
\begin{overpic}[width=.45\textwidth]{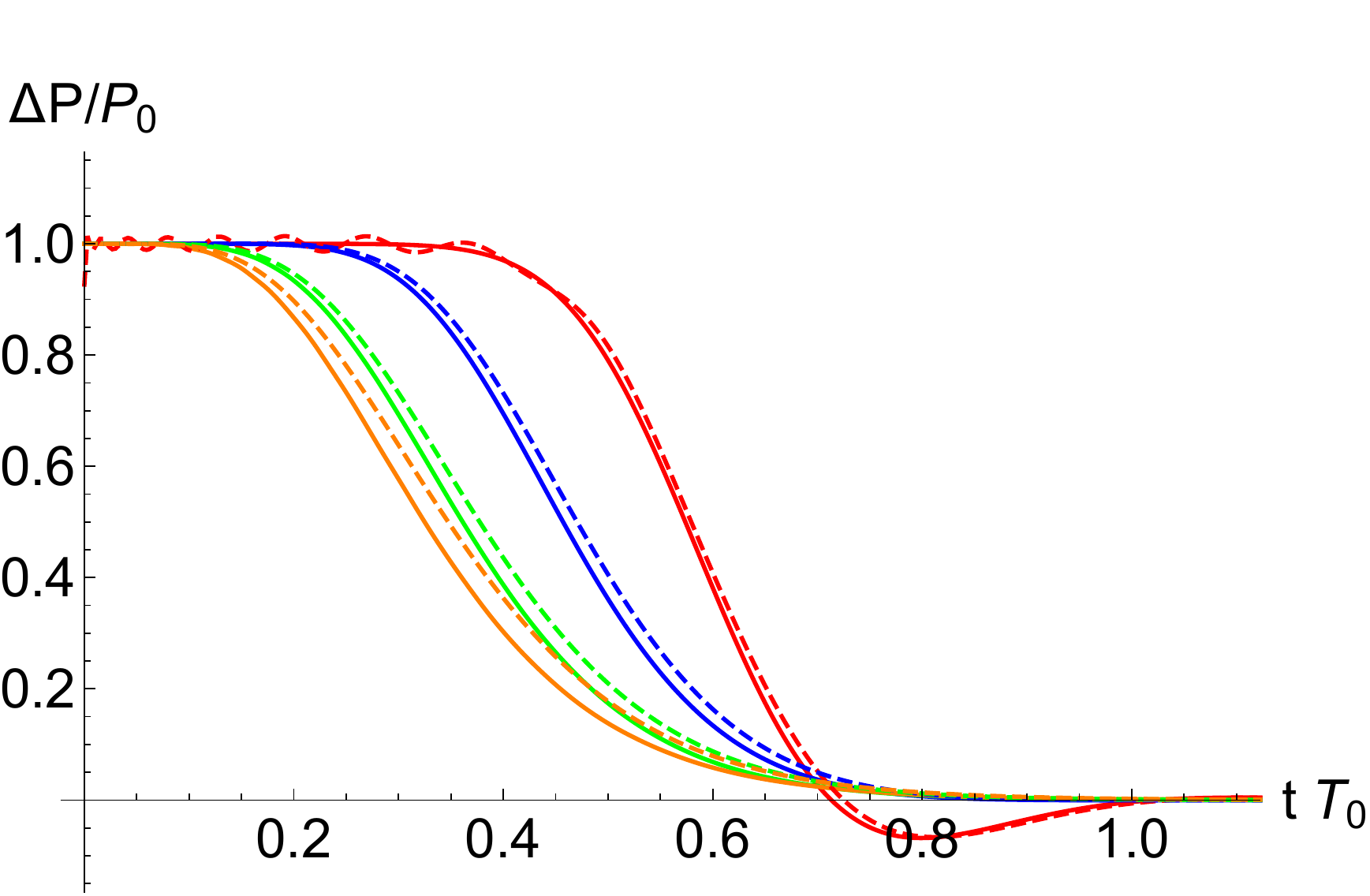}
		
	\put (45,25) {\includegraphics[scale=.28]{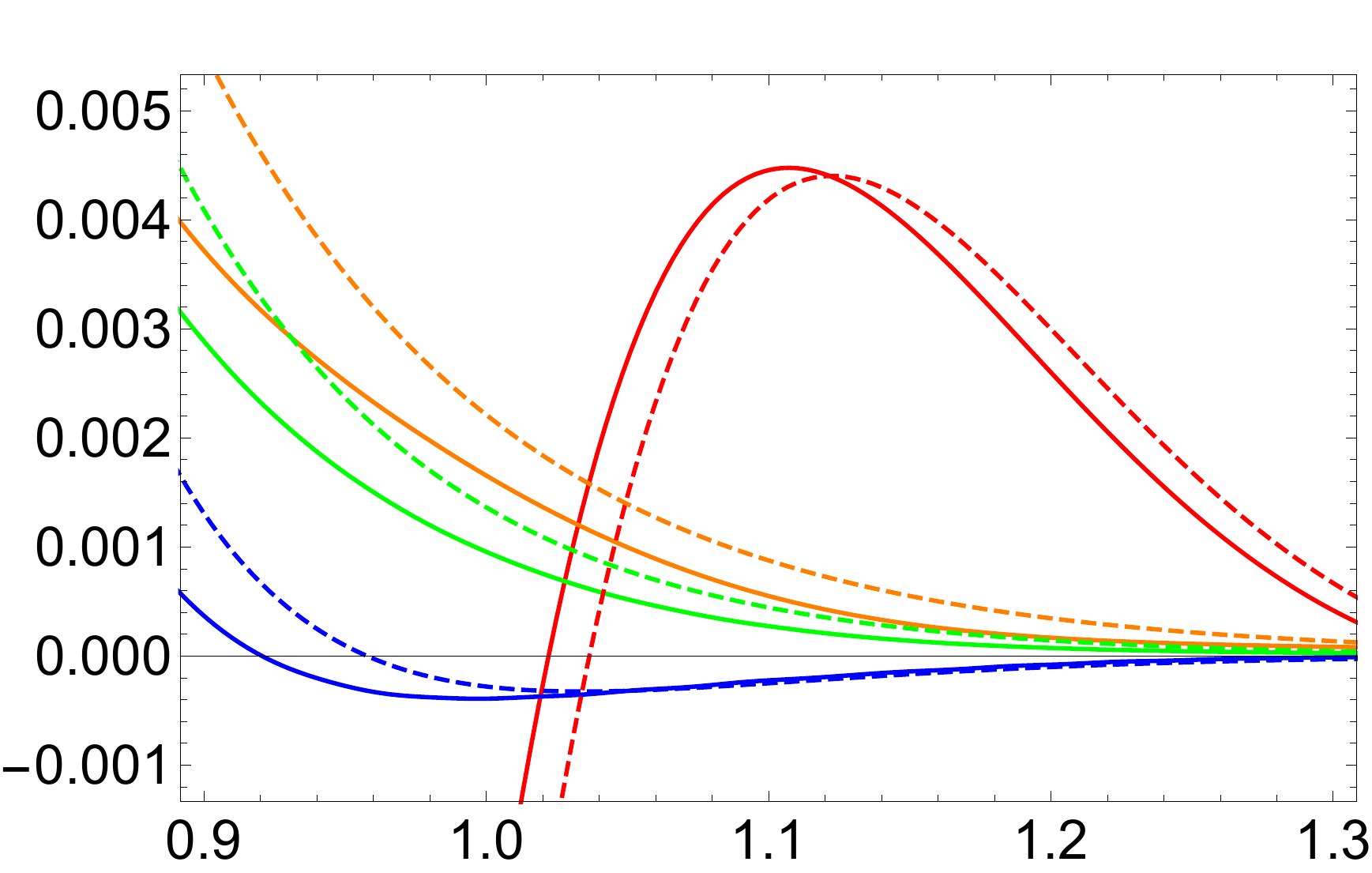}}
	\put (75,2) {\framebox(27,10){}}
	\put (102,2) {\rotatebox{82}{\noindent\rule{2.15cm}{0.4pt}}}
	\put (52.8,2) {\rotatebox{-230}{\noindent\rule{2.75cm}{0.4pt}}}
	
		\end{overpic}
\end{center}
    \captionsetup{justification=raggedright}
\caption{Evolution of the pressure difference, for $d=4$ and $z$ from 1 (rightmost) to 4 (leftmost). The blowup shows a small oscillation below 0 for $z=2$. Dashed lines are the evolution resulting from a fit of the first 10 quasinormal modes to the initial profile of $B$.}
\label{fig:Pevolution}
\end{figure}

The bulk evolution of $B$ describes the evolution of a pressure difference of the boundary theory through the function $\mathcal{P}$, which we turn to now.
In Figure \ref{fig:Pevolution} we show the time evolution of the pressure difference.
The system isotropizes on a timescale of the order of the inverse temperature.
Note first that for the cases where $z \geq d-1$, there are no oscillations, the system is overdamped.
This will be more clear when we look at the quasinormal modes, and is in agreement with \cite{sybesma:2015oha}.

The anisotropy $B$ obeys the massless Klein-Gordon equation at zero momentum. 
After redefining $\tilde{t} = z t$, this equation only depends on $d$ and $z$ through the ratio $\alpha \equiv z/(d-1)$, as 
\begin{equation}\label{eq:qnm}
\! \partial_u \left[ u^{- 1/\alpha} \partial_{\tilde{t}} B - u^{2-1/\alpha}  f(u) \partial_u B \right]\! +\! u^{- 1/\alpha} \partial_{\tilde{t}} \partial_u B \!= 0\ ,
\end{equation}
with $f(u)\!=\!u^{-2}(1\!-\!(u/u_{H})^{1+1/\alpha})$.
Note that in contrast, the nonlinear equations do depend on $d$ and $z$ separately.
To compute quasinormal modes we solve the associated generalised eigenvalue equation using pseudospectral methods \cite{Janik:2015waa}.

We compare the nonlinear evolution with the sum of the first 10 quasinormal modes, $B_{\text{QNM}} (u,t) = \text{Re} \sum_{i=0}^{9} c_i b_i(u) e^{- i \omega_i t}$. 
Here the $c_i$ are coefficients which we obtain by fitting to the initial profile $B(u,t=0)$, $b_i$ are the eigenmodes and $\omega_i$ the corresponding eigenfrequencies \cite{Heller:2013oxa}.
Dashed lines in Figure \ref{fig:Pevolution} show the fitted quasinormal mode evolution. They always lag behind on the nonlinear evolution a little, but otherwise give a very good approximation. 
The late time evolution of the pressure is well approximated by the lowest quasinormal mode.

\begin{figure}[h]
\begin{center}
\begin{overpic}[width=.45\textwidth]{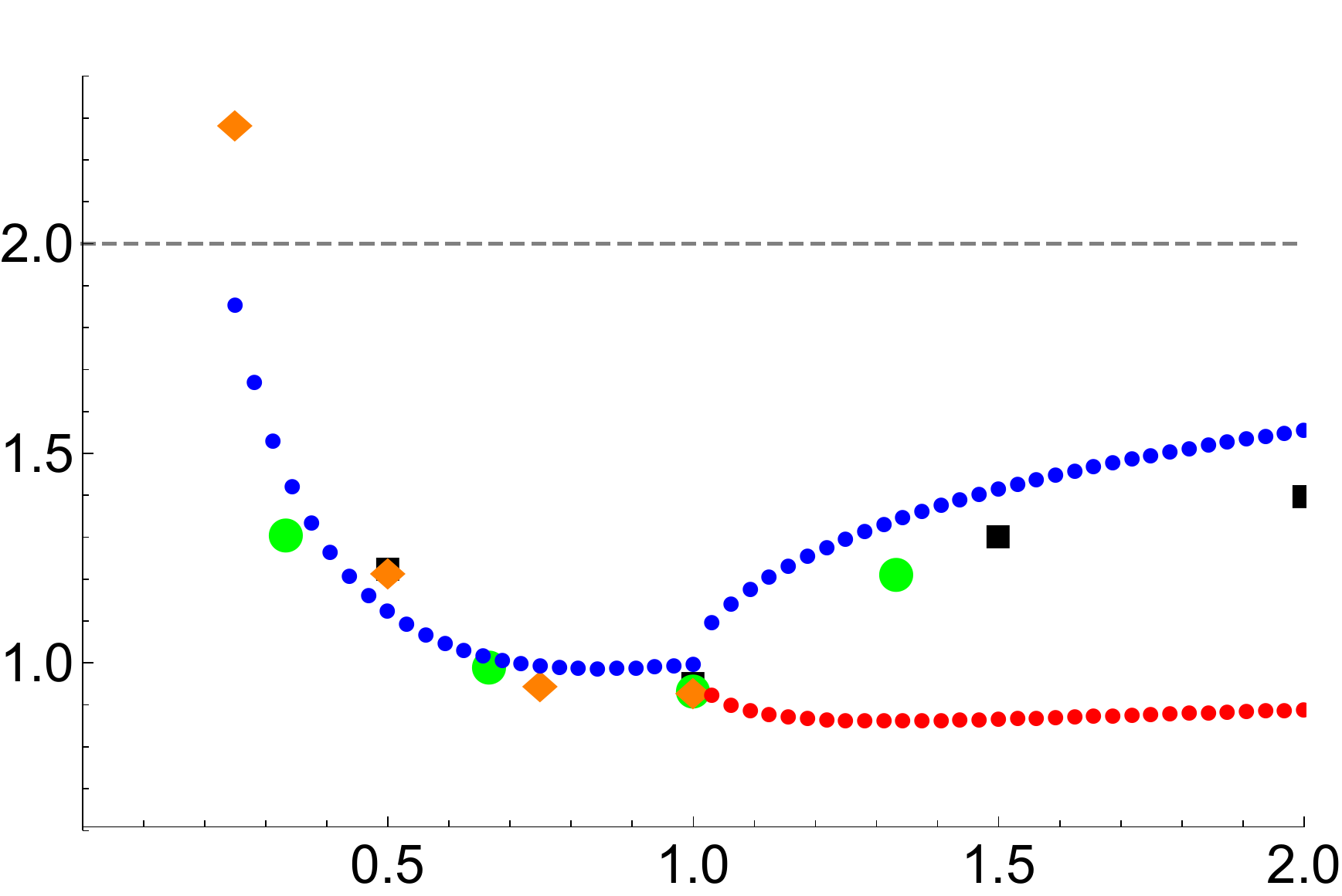}
 		\put (2,62) {$\displaystyle 4\pi T_{0}\tau$}
		\put (100,6) {$\displaystyle \alpha$}
		
		\end{overpic}
\end{center}
    \captionsetup{justification=raggedright}
\caption{Relaxation times from quasinormal modes (blue points) as a function of $\alpha \equiv \frac{z}{d-1}$.
At $\alpha=1$ the modes bifurcate, with the second lowest shown in red. 
Also shown are relaxation times $4 \pi T_0 \tilde{\tau}$ obtained from the nonlinear evolution, for $d=3$ (black squares), $d=4$ (green circles) and $d=5$ (orange diamonds).}
\label{fig:thermalization}
\end{figure}

\begin{figure}[h]
\begin{center}
\begin{overpic}[width=.40\textwidth]{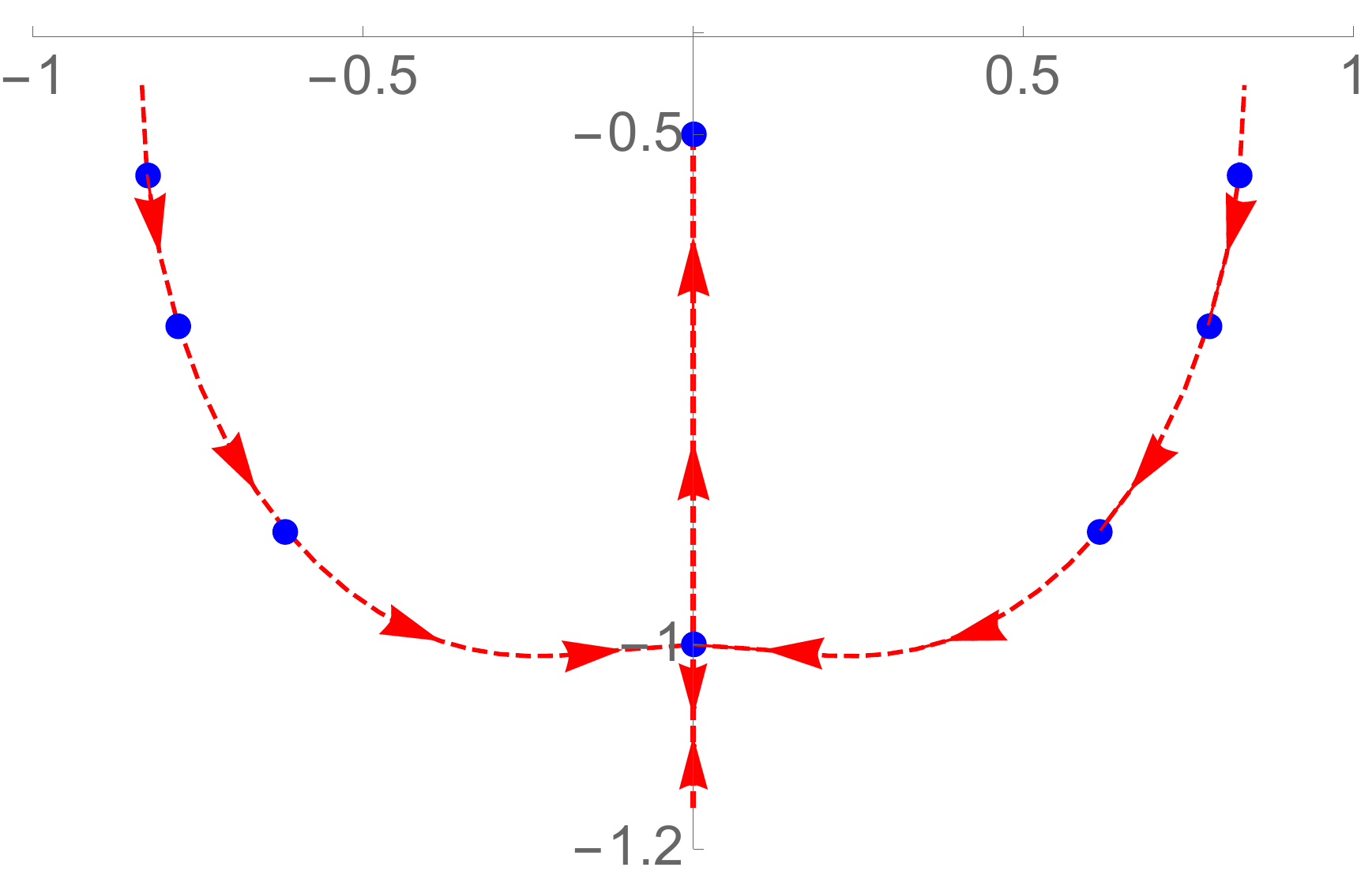}
 		\put (89.5,65.5) {$\displaystyle \frac{\omega_{\text{Re}}}{4\pi T}$}
		\put (53,2) {$\displaystyle\frac{\omega_{\text{Im}}}{4\pi T}$}
		
		\put (87,30) {$\displaystyle \alpha<1$}
		\put (4,30) {$\displaystyle \alpha<1$}
		
		\put (53,54) {$\displaystyle \alpha=\infty$}
		\put (53,12.5) {$\displaystyle \alpha=1$}
		
		\put (27,35) {$\displaystyle 1<\alpha<\infty$}

		\put (27,7.5) {$\displaystyle 1<\alpha<\infty$}

		\end{overpic}
\end{center}
    \captionsetup{justification=raggedright}
\caption{We display the lowest modes found from (\ref{eq:qnm}), using numerics. Following the flow of the arrows corresponds to increasing $\alpha$. For $\alpha>1$, one mode moves up and asymptotes towards $\omega_{Im}/(4\pi T)=-0.5$ as $\alpha\rightarrow\infty$. The other mode moves down, initially, but at some point reverses and for $\alpha\rightarrow \infty$ asymptotes towards the $\alpha=1$ location. Dots denote points with $\alpha$ corresponding to an AdS space.}
 \label{fig:movingmodes}
\end{figure}
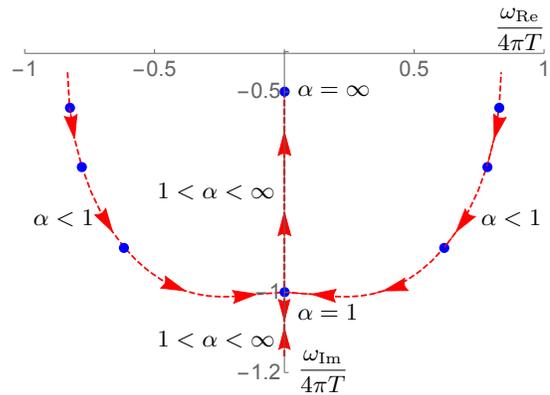

For $\alpha \geq 1$ the quasinormal modes become overdamped (purely imaginary) and we observe a bifurcation, see Figure \ref{fig:thermalization}, with one mode branching upwards and the other downwards, converging as $\alpha \rightarrow \infty$ to the first two modes in $AdS_2$. 
Notice $\alpha \rightarrow \infty$ can be interpreted as $z \rightarrow \infty$ at fixed $d$, or as $d \rightarrow 1$, at e.g. $z=1$, matching \cite{Hartnoll:2009ns} where it is stated that the $z\rightarrow \infty$ limit of Lifshitz corresponds to $AdS_2$. 
This behavior is  clarified in Figure \ref{fig:movingmodes} where we plot the motion of the lowest quasinormal modes in the complex plane, as we vary $\alpha$.

The value $\alpha = 1$ can be found from $AdS_3$, where one can show analytically that $\tau = \frac{1}{4 \pi T_0}$ \cite{Birmingham:2001pj,Myung:2012cb,sybesma:2015oha}. 
Curiously, this is not the minimal value of the relaxation time, which instead sits at $\alpha \approx 0.847227$ and reads $\tau\! \approx\! \frac{0.989002}{4 \pi T_0}$. Furthermore, for $\alpha = 1 + \epsilon^2$ with $\epsilon \ll 1$ we numerically find that $4 \pi T_0 \tau \approx 1 + \frac{1}{2} \epsilon$.

As $\alpha$ is taken to zero, the relaxation time diverges. However, we observe that with $d\! \leq \!5$, which covers all physically relevant cases and implies that $\alpha\geq\frac{1}{4}$, we have a range of relaxation times $0.989002\! \leq \!4 \pi T_0 \tau \!\leq \!2$.

From the nonlinear evolution we obtain an isotropization time, defined as $| \frac{\Delta P(t \geq t_{\text{iso}})}{P_0}  | \leq \epsilon$, for some choice of $\epsilon$. 
To compare this with the relaxation time from the quasinormal modes, we define $\tilde{\tau}\equiv t_{\text{iso}}/ \log(c_0 b_0(0)/\epsilon)$. 
When applied to a single quasinormal mode decaying as $B_0(t) = c_0 b_0(0) e^{-t/\tau}$ , $\tilde{\tau}$ becomes equivalent to $\tau$.

Along with the relaxation times in Figure \ref{fig:thermalization} we plot $4 \pi T_0 \tilde{\tau}$, with $c_0$ and $b_0(0)$ obtained from the quasinormal mode fit to the nonlinear evolution, taking $\epsilon = 0.1$. We checked that choosing a smaller $\epsilon$ brings the nonlinear evolution closer to the evolution dictated by the quasinormal modes, as expected, since smaller $\epsilon$ corresponds to later times.    
These have qualitatively the same dependence on $\alpha$, but generally lie just below $\tau$ due to the presence of higher order modes. There can be some noise due to oscillations, which are not accounted for in the definition of $\tilde{\tau}$, as seen in the point at $\alpha = 1/4$.
Note also that at $\alpha = 1$ we have the three cases $(d,z) = (3,2)$, $(4,3)$ and $(5,4)$ which are very close to each other in the plot, and for $\alpha = 1/2$ the cases $(d,z) = (3,1)$ and $(5,2)$ nearly overlap.

\paragraph{Discussion.---} 
Our main result is the holographic computation of the isotropization time of the pressure difference in a nonrelativistic plasma, taking into account full back reaction. 
We also confirm the existence of the overdamped regime for $d\leq z+1$ which was predicted by \cite{sybesma:2015oha} using quasinormal modes . 
Moreover, we find that for physically relevant dimensions the equilibration time is roughly of the order $1/(4\pi T_{0})$, regardless of the value of $z$. We note that this universal behavior only applies to critical theories with no mass gap. 
Holographic equilibration in gapped theories such as QCD has also been studied in the literature \cite{Ishii:2015gia,Buchel:2015saa,Janik:2015waa}  where the approach to equilibrium may be qualitatively different \cite{Ishii:2015gia}.

Our results are valid for large $N^2\sim 1/G$ and strong coupling, but are consistent with the general arguments about equilibration times for general strongly coupled quantum critical systems (see e.g. \cite{sachdev2007quantum}). This seems to indicate that our results are quite robust and the large $N$-limit does not affect this behavior qualitatively.

We note that the perturbations we take are large, i.e. $\Delta P\sim P_0$, yet the backreaction is quite small, i.e. the entropy increase is only a few percent. 
It might be interesting to consider larger profiles at fixed $\alpha$ but different $d$ and $z$, to see which combination will reach equilibrium faster deeper into the nonlinear regime.
It has proven difficult to do this with significantly larger profiles, but using domain decomposition \cite{boyd2001chebyshev} might help isolate the numerical difficulties at the boundary. 
%

\paragraph{Acknowledgements.---}

We thank Takaaki Ishii and Wilke van der Schee for comments on the draft. This work was supported by the Netherlands Organisation for Scientific Research (NWO) under VIDI grant 680-47-518 and VICI grant 680-47-603, and the Delta-Institute for Theoretical Physics (D-ITP) that is funded by the Dutch Ministry of Education, Culture and Science (OCW).

\bibliography{bibliography}

\end{document}